\documentclass[preprint2]{emulateapj}
\usepackage{epsfig}

\def\jfm{{\rm J. Fluid Mech.}\,}

\def\mearth{M_\oplus}
\def\cm2s{{\rm cm}^2\,{\rm s}^{-1}}
\def\ga{\,\hbox{\hbox{$ > $}\kern -0.8em \lower 1.0ex\hbox{$\sim$}}\,}
\def\la{\,\hbox{\hbox{$ < $}\kern -0.8em \lower 1.0ex\hbox{$\sim$}}\,}

\begin{document} 

\title{Heat transport in giant (exo)planets: a new perspective} 
\shorttitle{Heat transport in giant planets: new perspective}

\author{Gilles Chabrier and Isabelle Baraffe \altaffilmark{1}}
\affil{Ecole Normale Sup\'erieure de Lyon, CRAL (UMR CNRS 5574), Universit\'e de Lyon, France}

\altaffiltext{1}{and Max Planck Institut for Astrophysics, Garching, Germany}

\newpage

\begin{abstract}
We explore the possibility that large-scale convection be inhibited over some regions of
giant planet interiors, as a consequence of a gradient of composition
inherited either from their formation history or from particular events like giant impacts or core erosion
during their evolution. Under appropriate circumstances, the redistribution of the gradient of molecular weight can
lead to double diffusive layered or overstable convection. This leads to much less efficient heat transport
and compositional mixing than large-scale adiabatic convection.
We show that this process can explain the abnormally large radius of
the transit planet HD209458b and similar objects, and may be at play in some giant planets,
with short-period planets offering the most favorable conditions. Observational signatures of
this transport mechanism are a large radius and a reduced heat flux output compared with
uniformly mixed objects.
If our suggestion is
correct, it bears major consequences on our understanding of giant planet formation, structure and
evolution, including possibly our own jovian planets.
 \end{abstract}
 

\newpage

\section{Introduction}

Over 200 extrasolar giant planets have now been discovered by radial velocity surveys.
Fourteen of these planets have been observed transiting their parent star,
allowing an accurate determination of their radius and mean density.
For half of these objects, notably the first ever discovered transit HD209458b,
the predicted theoretical radius lies several $\sigma$'s (up to $\sim 20\%$) below the observed mean value \citep{B05}. 
The various scenarios proposed so far to solve this discrepancy either have been rejected on
observational or theoretical arguments \citep{Levrard07} or lack an
identified robust mechanism to convert surface kinetic energy
into thermal energy at depth \citep{SG02}.
A fair conclusion of these studies is that an important physical mechanism is probably missing in our present description of at least some short
period planets, for which we have a radius determination, but possibly of all extrasolar or even solar
giant planets.

According to the conventional core-accretion model for planet formation \cite{P96},
planets are believed to have a substantial enrichment in heavy elements
compared with their parent star, with a total of $\sim 40\mearth$ for a Jupiter-mass object \cite{Alibert05}.
Observational constraints of Jupiter and Saturn show that these planets do have a significantly enhanced
Z-abundance compared with the Sun, with a global mean mass fraction $Z$=$M_Z/M_p\approx$ 10-20\% \cite{SG04}.
In all these calculations, the big planetesimals are generally supposed to drown to the
core during the early phase of solid accretion, while the smaller ones are distributed {\it uniformly} throughout the envelope,
leading to a uniform heavy element abundance.
This is a rather simplistic description of the planet internal structure, which implies (i) well defined
interfaces between the central core and the (highly diffusive) H/He rich envelope and (ii) very efficient
large-scale thermal convection
throughout the entire gaseous envelope. The observed
atmospheric abundances of Jupiter and Saturn, however, seem to require the redistribution of a subsequent
fraction of heavy elements in the interior of these planets \cite{Guillot04}.
 In this Letter, we explore the consequences of the presence of
an initial compositional gradient in the envelope, as a result of either early planetesimal accretion
or subsequent core erosion, and of the resulting less efficient heat transport and
compositional mixing on the fate of gaseous planets. We show that layered 
convection, if it occurs
as a result of this compositional gradient,
might be the lacking physical mechanism to explain the transiting planet abnormally large
radii.

\section{Layered semiconvection}

The  presence of a positive compositional gradient, i.e. a gradient of mean
molecular weight $\nabla_\mu$=${d\ln \mu\over d\ln P} >0$, tends to stabilize the fluid
against convective instability according to the Ledoux stability condition: 

\begin{eqnarray}
\nabla_{ad}>\nabla_T+ {\chi_{\mu}\over \chi_T}\nabla_\mu \, ,
\label{Ledoux}
\end{eqnarray}

\noindent where $\nabla_T$ and $\nabla_{ad}$ denote the usual temperature and adiabatic gradients,
respectively, and $\chi_\mu$=$({\partial \ln P\over \partial \ln \mu})_{\rho,T}$, $\chi_T$=$({\partial \ln P\over \partial \ln T})_{\rho,\mu}$.
In most of the giant planet interiors, superadiabaticity  is extremely small, with $\nabla_T-\nabla_{ad}\lesssim 10^{-8}$, so a small molecular weight gradient over a typical mixing length size region can affect significantly and even damp out convection.
In convective systems where buoyancy effects of (destabilizing) heat and (stabilizing) composition are opposed, the process leads generally to quasi-static uniformly mixed convective layers separated by small diffusive interfaces with steep gradients, $\nabla_\mu \gg 1$. 
Such stable layered convection is indeed observed in some
areas of the Earth's oceans, due to the presence of the stabilizing salt gradient (thermohaline convection)
leading to a stratified step-like temperature profile with stable boundary layers \cite{Schmitt}.
Laboratory experiments have also confirmed this layering \cite{Fernando89}.
Once formed, the stratification is stable provided the compositional
gradient remains large enough at the interfaces to satisfy Eq.(\ref{Ledoux}). 
Would such a stratification occur in planetary interiors,
the layered part of the interior can be considered as a {\it semiconvection} zone with a reduced
efficiency to transport the internal heat and composition flux compared with large-scale convection.
One may argue that the conditions in planetary interiors differ from the ones in the oceans or in
the experiments.
Characteristic thermal diffusivity in H/He planetary interiors, dominated by electronic transport in the
central, ionized parts, and by molecular motions in the outer envelope, lies in the range $\kappa_T\approx 10^{-1}$-$10^{-2}$ $\cm2s$, while the
kinematic viscosity is $\nu\approx 10^{-3}$-$10^{-2}$ $\cm2s$ \cite{SS77}. The characteristic Prandtl number thus ranges from $Pr=$ $\nu/\kappa_T\approx 10^{-2}$ to 1.
These values do not differ by large factors from the ones characteristic in the oceans or in
laboratory experiments, $Pr\approx 1$-10, in contrast
to the ones characteristic of stellar conditions ($Pr < 10^{-6}$). Therefore, a
layering process in giant planet interiors can not be excluded and it is worth exploring the
consequences on the planet evolution. Such layered convection may occur near the
discontinuity in composition at the boundary of the central rocky-icy core or in chemically
inhomogeneous regions in the interior, reminiscent of the early planetesimal accretion
episodes.

There is presently no widely accepted description of semiconvection. Water-salt experiments \cite{Fernando89}
show that a series of quasi-static convective layers separated by
diffusive interfaces develop when a balance is reached between the variation of potential energy (i.e. of buoyancy)
due to mixing at the interface and the kinetic energy of the
eddies available at the interface. This translates into a critical Richardson number $Ri$=$l\,g\Delta \rho/\rho \langle v^2\rangle$ of order 1 to 10, where  $\Delta \rho \simeq \rho (\Delta T/ T)$ is the density contrast between the diffusive
and convective layers,
$g$ is the gravity and $\rho \langle v^2\rangle$ is the kinetic energy of the convective flow, of characteristic average length scale $l$ and rms velocity $\sqrt{\langle v^2\rangle}$. 
Guided by experimental results \cite{LS78}
and energetic arguments,
Stevenson (1979), in a wave description of semiconvection, showed that, should layers form as a result of small-scale wave breaking whereupon the compositional gradient is redistributed,
they would be stable if

\begin{eqnarray}
(\kappa_T+\nu)(\nabla_T-\nabla_{ad})<(D+\nu)\nabla_\mu,\,\,{\it i.e.}\,\,Pr \ga \tau^{1/2}
\end{eqnarray}

\noindent where the inverse Lewis number $\tau$=$Le^{-1}$=$D/\kappa_T$ is the ratio of
the solute microscopic diffusivity to the thermal diffusivity. Spruit's (1992) stability condition is less restrictive,
since layered formation is supposed to always occur, and is given essentially by Eq.(1).
Under jovian planet conditions, typical values
are $D\approx 10^{-3}$-$10^{-4}$ $\cm2s$ \cite{SS77}, then $\tau \sim 10^{-2}$, so that, according to this criterion, diffusive
layers could be at least marginally stable in giant planet interiors. 
It is worth noting that the molecular to thermal diffusivity ratio is the same for a H/He mixture under jovian interior conditions
as for salty water, $\tau\approx 10^{-2}$, so that the extent of the solute versus thermal layer is about the same. Since, according to experiments and condition (2), this ratio is the relevant criterion for stability of the layers, this adds some support to the planetary case. 

In a layered convection stratification, heat is carried away from the interfaces by descending and ascending plumes in the overturning regions while 
transport across the interface occurs by diffusion. 
Because of the boundary layers, only a part of the fluid transports heat efficiently.
On average, one has "convective-like" motions having a much shorter length scale than for ordinary convection.
The {\it thermal} thickness of the diffusive layers, $\delta_T$, is determined by a balance between the
thickening due to diffusion and the entrainment due to convective motions so 
the convective time, $\sim l/v$, in the mixed layer of size $l$ must be comparable to the thermal diffusion time, $\sim \delta_T^2/\kappa_T$, across the boundary layer \cite{Fernando89}. 
This yields $\delta_T \approx (\kappa_T\,l/v)^{1/2}$ and $\delta_X \approx (D\,l/v)^{1/2}$ for the thickness of the heat and compositional interfacial layers, respectively. 
The number $N$ of layers is of course very uncertain.  
It can be crudely estimated as follows. The heat flux $f$ transported by convection in each mixed layer is the mass flux carried by the plumes fed from the diffusive layers, and thus of width comparable to these layer thickness, times the energy variation across the convective layer, $T\delta S\approx T(l\,\nabla S)$ :

\begin{eqnarray}
f={1\over l}(\rho v \delta_T)\times (T \delta S)
=\rho c_p \kappa_{T} \frac{l}{\delta_T}{T\over H_P}(\nabla - \nabla_{ad})
\end{eqnarray}

\noindent For a semiconvective region extending over a planet-size region, the total number of layers is thus given by:

\begin{eqnarray}
N\approx {F_{tot}\over f}={L/4\pi R^2 \over \rho c_p \kappa_{T} {T\over H_P}(\nabla - \nabla_{ad})}\times
\frac{\delta_T}{l}
\end{eqnarray}

\noindent Using characteristic numbers for jovian planet conditions (with $l\sim H_P$),
one gets
$N\sim 10^5$-$10^6$ as a rough estimate. If convection is inhibited, however,
the smaller heat flux and larger superadiabaticity require less layers.

\section{Effect on the evolution}

We have conducted calculations following the evolution of a template Jupiter-mass planet,
representative of HD209458b and similar short-period planets, with a global
metal content $M_Z=40\,\mearth$, including a 6 $\mearth$ core, i.e. $M_Z/M_p=13\%$ ($Z\simeq 6\,Z_\odot$), in agreement with previously mentioned planet formation models
and Jupiter and Saturn's observational constraints. The amount of heavy elements is distributed
initially throughout the
planet following a gradient, distributed within a certain
number of boundary layers $N$ where condition (1) is fulfilled. The layers are
located within the inner $\sim 30\%$ by mass (60\% in radius) of the planet, where H and He are fully
ionized, to ensure high enough thermal conductivity.
The present calculations have been done with $N$=50 and $N$=100; the width of
each boundary layer corresponds to $\delta_T\approx 10^3$ cm $\sim 10^{-6}H_P$. 
A larger number of layers would be computationally too difficult to resolve correctly.
These boundary layers
are separated by larger convective, mixed layers, with a uniform composition ($\nabla \mu=0$), where
the usual mixing-length formalism applies.
The sizes of the boundary and mixed layers ($\delta_T, \delta_X, l$) obey the aforementioned relationships. 
The heat flux $F_T$ and solute flux $F_Z$ in the boundary
layers are calculated with the appropriate diffusion equations: $F_T$=$\rho c_p \kappa_T \nabla T$, where $\kappa_T$=$\frac{16 \sigma T^3}{3\rho^2 c_p}
(\kappa_c^{-1}+\kappa_r^{-1})$ and $\kappa_c, \kappa_r$ denote the conductive \cite{P99} and
radiative \cite{OPAL,Ferguson05} mean opacities, respectively, and $F_Z$=$\rho D \nabla Z$, where $\nabla Z
$=$-{Z\over \chi}H_P^{-1}\nabla_\mu$, with $\chi= ({\partial \ln \mu \over \partial \ln Z})$,
is the mean concentration gradient across the layer. 
Conduction remains efficient enough in the thin boundary layers to fulfill condition (1). Because diffusion limits the heat transport, the internal heat flow of the planet is significantly reduced compared with that of a fully
convective object. The signatures of double-diffusive convection in a planetary interior are thus a reduced heat output
and a larger radius compared with an object where heat is transported efficiently by large-scale
convection. This is illustrated in Fig. 1, which compares the evolution of the radius and
thermal intrinsic luminosity of the planet in both cases. The excellent agreement with the otherwise unexplained
observed radii of HD209458b and 
similar irradiated planets suggests that diffusive convection might be taking
place in the interior of at least certain giant planets. As seen, the expected luminosity
at young ages is more than one order of magnitude fainter than that of a fully convective planet evolving
from a comparable initial state.
The observational confirmation
of the present scenario would be either the determination of an exoplanet temperature or luminosity
at young ages\footnote{For short-period, irradiated planets, however, the intrinsic flux of the planet, $\sigma T_{eff}^4$, is smaller
than the absorbed and reflected contributions of the incident stellar flux, $\sim (R_\star/a)^2{\mathcal F_\star}$. For long-period planets or for planed telescopes like the LBT or the JWST, dedicated to infrared planet searches, however, the planet intrinsic luminosity can be determined.}
or the observation of an inflated radius for a transiting planet at large enough orbital distance, $a\ga 0.1$ AU for a solar-type parent star, for stellar irradiation not to affect the planet's internal structure.
Figure 1 illustrates also the dependence of the evolution upon the number of layers. Less boundary
layers implies larger convective layers and thus more efficient heat transport, as illustrated by the more rapidly decreasing radius in the 50-layer calculations.

\begin{figure}
\epsscale{1.1}
\plotone{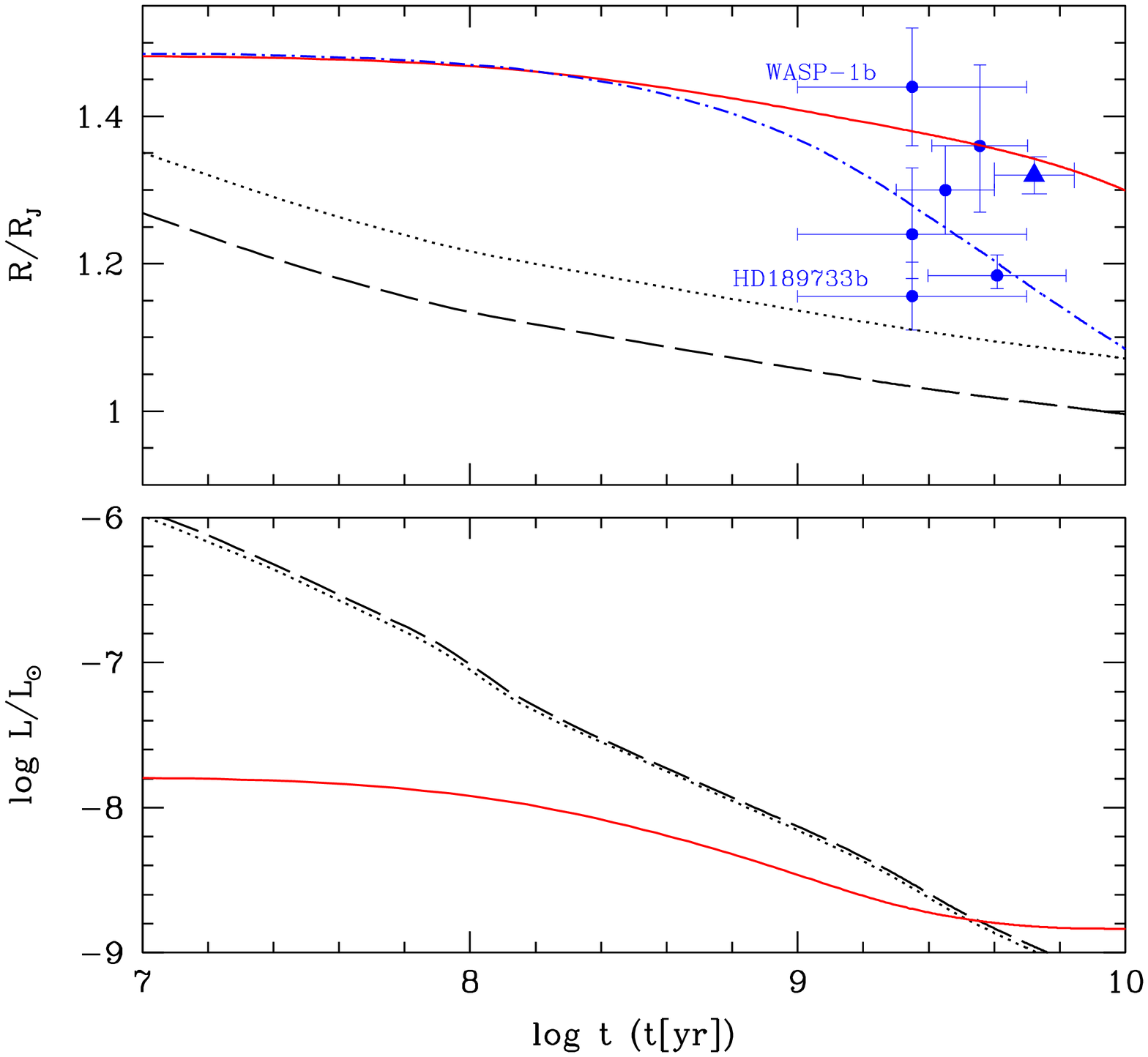} 
\caption{{\footnotesize Evolution of the radius and the intrinsic luminosity of a Jupiter-mass ($1\,M_{Jup}$) planet orbiting at 0.05 AU of a Sun-like star, from the same initial
conditions. All calculations include the effect of the stellar irradiation on the planet structure and evolution. (1) Effect of a core and metal enrichment: dotted line: adiabatic interior, no core, $Z=Z_\odot$; long-dashed line : adiabatic interior, central dunite core
$M_c=6\,\mearth$, metal enrichment $Z_{env}\simeq 6 \times Z_\odot$ in the envelope. Effect of layered convection (same $M_c$ and $Z_{env}$):
solid line: layered convection for $N$=100 layers; dot-dash line (upper panel): $N$=50 layers. The observed values of
 the seven planets with abnormally large radii, HD209458b (triangle), WASP-1b, XO-1b, OGLE-TR-56b, Tres-2,
HAT-P-1b and HD189733b, with masses ranging from 0.5 to 1.3 $M_{Jup}$, are displayed with their most
recent 1-$\sigma$ error bar determinations.}}
\end{figure}

A key question is to know if diffusive interfaces can persist on time scales comparable to the characteristic time for the evolution of the planet. 
According to the aforementioned critical Richardson number criterion, supported by experiments \cite{Fernando89}, a quantitative argument is
that if the average kinetic energy in the convective layers
is smaller than (a fraction of) the potential energy wall of the
interface, convection cannot penetrate deeply into this latter and significant entrainment across the interface cannot occur. This implies that the molecular
diffusion time scale be long enough.
This latter can be estimated for the entire stack of layers, distributed over a region of size $L$ in the planet (presently $L\sim 10^9$ cm). The flux of element across an interface is $F_Z\approx \rho D\frac{\delta Z}{\delta_X}$, where
$\delta Z\approx {l\over L}\,\Delta Z $ is the jump in the element mass fraction at each interface while $\Delta Z$ is
the total variation over the entire semiconvective region. The time scale to redistribute the entire gradient over the
entire region is then $t\approx {\rho L\Delta Z\over F_Z}\approx \frac{L^2}{D}\frac{\delta_X}{l}
\approx $ 10 Gyr. This admitedly crude estimate shows that the stable diffusive convection configuration might last long enough to affect substantially the evolution. With the typical value $D=10^{-3}\cm2s$, about 10\% of the initial gradient $\Delta Z$ has been transported by diffusion over a Gyr,
as confirmed by our numerical calculations. In principle, the
compositional gradient thus remains large enough during the
evolution for the Ledoux criterion to remain valid in a majority of layers. In other words, the temperature jump at interfaces
is too small to offset the molecular weight stabilization of
interfaces ($\frac{\Delta T}{T} \la \frac{\Delta \mu}{\mu}$). The composition and temperature profiles in our calculations
at 5 Gyr are portrayed in Fig. 2, with $\delta Z/\Delta Z\approx 1\%$ and $\Delta T\approx 10^3$ K at each diffusive interface.
Note that, if layers form in sequence through turbulent entrainment or
from sporadically breaking internal waves generated by oscillatory instabilities, interfaces may be
dynamically renewed with time, if some compositional gradient or stirring effects remain present.
Such a process occurs in laboratory systems and oceans.

\begin{figure}
\epsscale{0.9}
\plotone{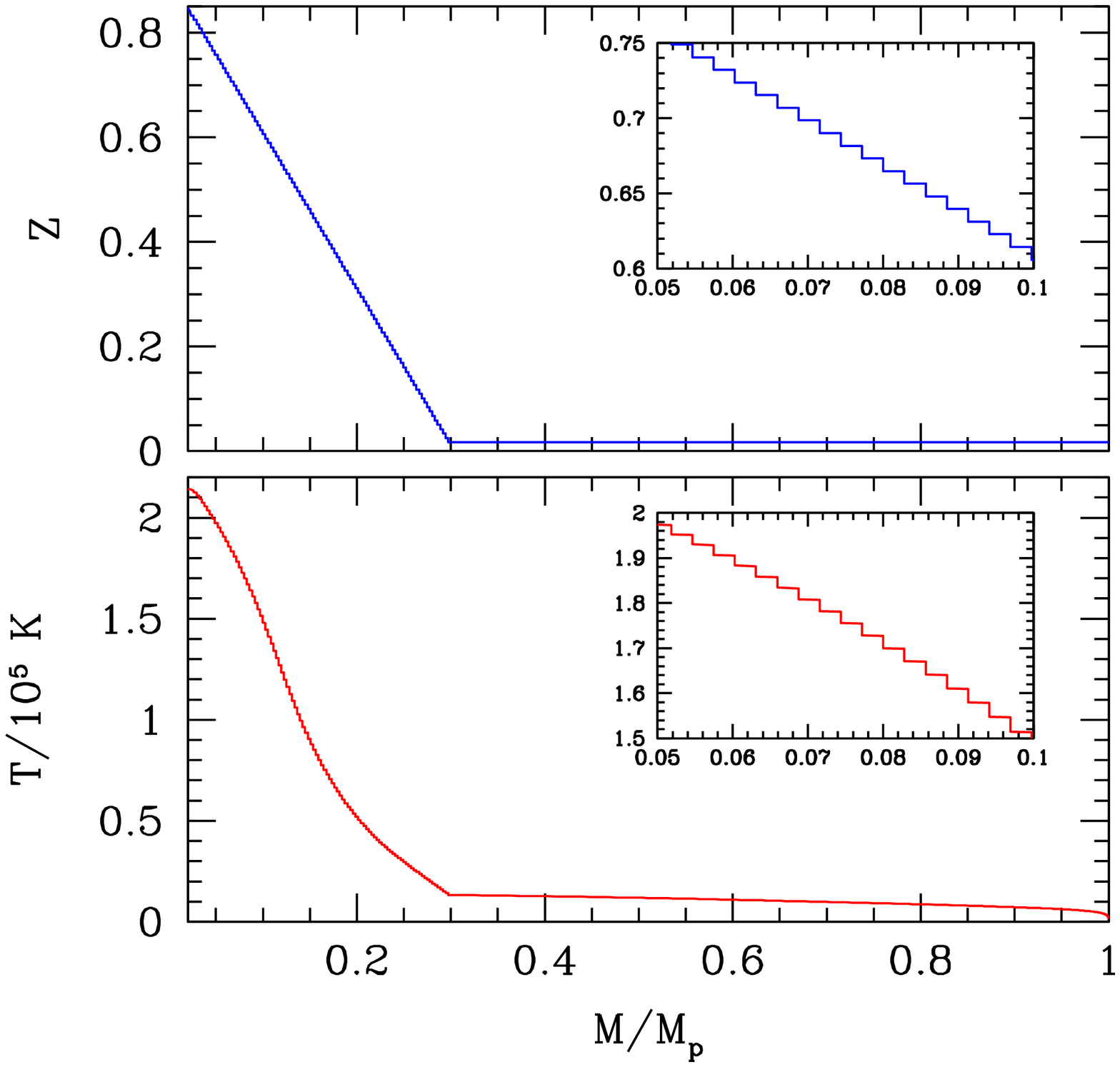} 
\caption{Internal heavy element and thermal profiles at 5 Gyr for the case of 100 convective + diffusive
layers distributed in the inner 30\% by mass of the planet. The global heavy element mass fraction of the planet is $Z=M_Z/M_p=0.13$, including the central core ($M_{core}=2\%\,M_p$). 
Note the step-like T and Z-distributions, as portrayed in the inner subsets.}
\end{figure}

Different reasons can be advocated for the cause of the initial compositional gradient. This latter can be inherited from the formation process. Large incoming planetesimals could disseminate part of their constituents,
iron, silicates, ices, by ablation and break-up as they
penetrate the building gaseous envelope \cite{IaroslavitzPodolak07}. Note also that accretion will not
proceed homogeneously as capture mechanisms differ for the gas (H, He), ice (essentially C, O, N) and rock (silicates
and iron) components.
This will increase substantially the compositional gradients before the core is reached. As mentioned
earlier, even modest gradients
can easily offset superadiabatic excess over planet-size regions, preventing large-scale convective
motions. 
A compositional gradient might also result from disruption and redistribution of the core due to a giant impact or
erosion at the core-envelope
interface because of metallic hydrogen high diffusivity, 
leading to a core diluted into a fraction of the planet \cite{Stevenson82,Guillot04}. 
The redistribution of these elements might be
partially inhibited by diffusive processes, forming diffusive interfaces because of the opposite buoyancy effects of heat and composition.
Furthermore, when accreting the envelope, 
part of the outermost regions of the protoplanet might be nearly isothermal \cite{Mizuno80}, which favors the stability of a compositional gradient.
Interestingly, when distributing the layers in the {\it outer} 10\% by mass ($\sim 15\%$ by radius)
of the planet, where {\it radiative} thermal diffusivity starts to dominate over conductive diffusivity,
we get an effect
similar to the one portrayed in Fig. 1. 
The situation for the formation of diffusive convection is particularly favorable for short-period exoplanets
for several reasons. First of all, a substantial fraction of the gaseous envelope has been
eliminated by evaporation \cite{B06}, leading to a larger metal fraction.
Second of all, for short-period exoplanets, the numerous collisions tend to eject the gas, leading
to a larger enrichment in
planetesimals than for the other planets. 
Third of all, the higher internal temperatures for short-period, irradiated planets, favor
(i) ionization of the various elements and thus the thermal conductivity, (ii) solubility of the core material into the envelope. At last, because of the stellar irradiation,
the outer layers of short-period planets are isothermal and not adiabatic.

\section{Conclusion and perspective}

The aim of the present Letter is to suggest an alternative,
possibly important energy transport mechanism in giant planet interiors and to explore its effects
on the evolution. These calculations provide a consistent description of the evolution of
giant planets, {\it with a metal enrichment in agreement with observational constraints}, in case heat is transported by layered convection. Assuming an initial compositional stratification within a certain
number of double-diffusive interfaces,
diffusive and convective transport in the respective layers are calculated consistently during the evolution. Only the outermost 40\% in radius (70\% in mass) of the planet can convect freely.
These calculations, however, cannot be expected to give an
accurate description of the onset and stability of layered convection. There is presently no
accurate treatment of this mechanism under conditions characteristic of giant planets.
The only attempt to study the onset of double-diffusive layer formation at low Prandtl numbers ($Pr<1$)
\cite{Merryfield95} remains inconclusive. 
Indeed, insufficient numerical resolution and artificially enhanced viscous and molecular diffusivities
in the simulations might suppress small-scale motions/instabilities and the formation of a statistically steady state of intermitent diffusive layers. 
Simulations of vertical salinity in water, $Pr$=$7$, on the other hand, well reproduce the experiments 
and confirm the formation of quasi-static convective
layers separated by diffusive interfaces above a critical Richardson number \cite{MolemakerDijkstra97}. 

Even though the present calculations rely on some uncertain ground,
the agreement with the puzzling and otherwise unexplained observed radius of HD209458b
and other abnormally large exoplanets leads to the
conclusion that layered
convection might be taking place in at least some planets and could explain
their particular properties.
This transport mechanism yields a much reduced heat escaping rate compared with a homogeneous adiabatic structure.
Even if a stable layered configuration does not occur,
overstable modes of convection (fullfilling Eq.(1) but not Eq.(2)), due to the presence of opposite diffusive processes (composition and heat) of different efficiencies,
can lead to the growth of small-scale fluid oscillations \cite{SS77b, Stevenson79}.
Overstability, however, is more similar to an enhanced diffusion process
than to a convective mixing process, with a much smaller energy transport efficiency, as shown
by experiments \cite{SS77b}.
The onset or persistence of layered or overstable (oscillatory) convection
might require optimal conditions, inherited from particularly
favorable formation or evolution histories (e.g. late accretion of large unmixed planetesimals or
giant impacts stirring up completely the planet interior).
The present paper suggests that, under such appropriate conditions, the heat transport mechanism in giant planet interiors can be severely affected, decreasing the efficiency of or even inhibiting large-scale convection. This should motivate 3D investigations of convection in the presence of a stabilizing compositional gradient under conditions suitable to giant planets and the search for transits at larger orbital distances.

\acknowledgments
The authors are indebted to the MPA
for the
warm hospitality. We are grateful to Henk Spruit for valuable discussions and to Friedrich Kupka
for pointing out useful references.

\clearpage

\begin{figure}
\epsscale{1.0}
\plotone{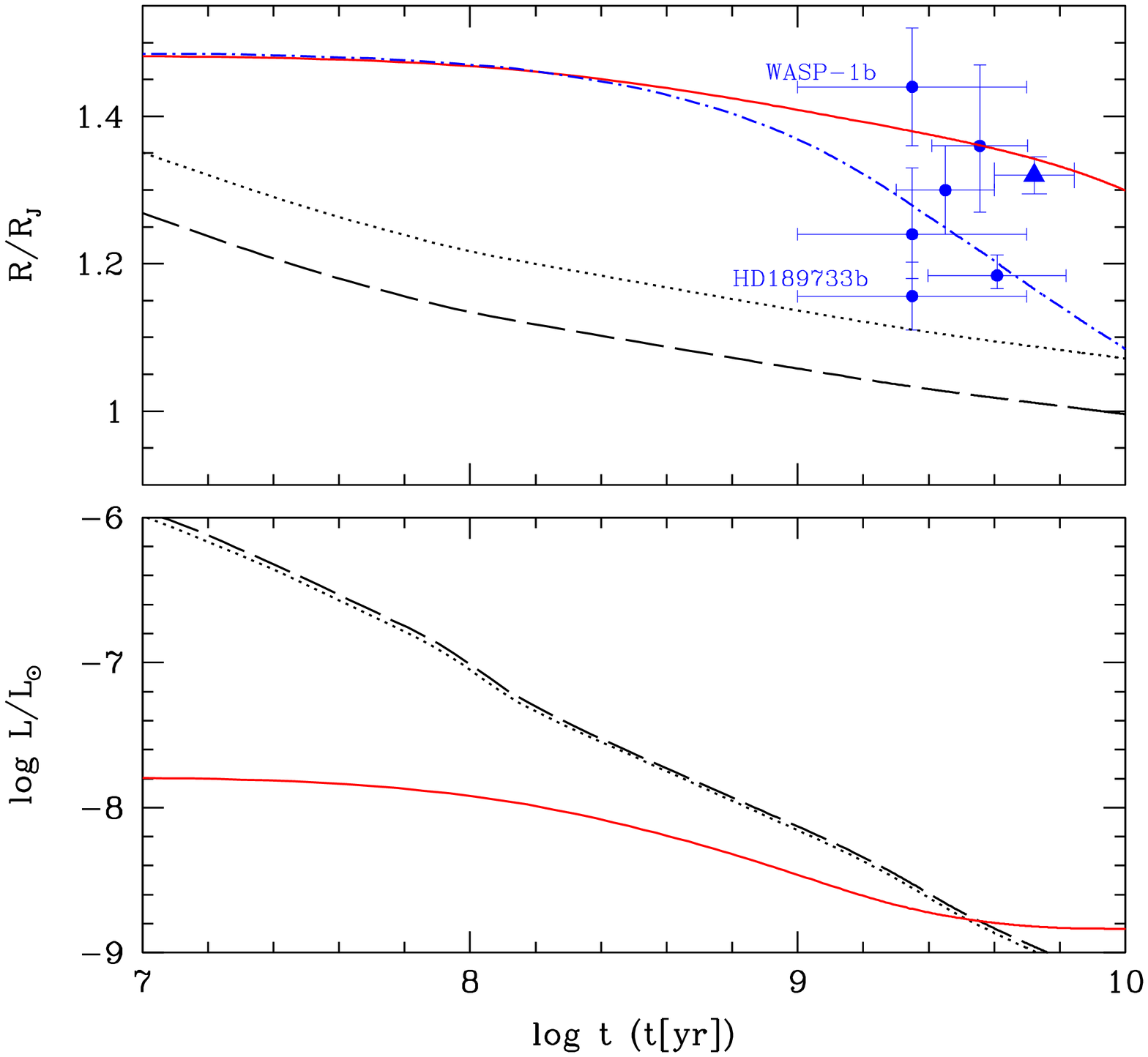} 
\caption{{\footnotesize Evolution of the radius and the intrinsic luminosity of a Jupiter-mass ($1\,M_{Jup}$) planet orbiting at 0.05 AU of a Sun-like star, from the same initial
conditions. All calculations include the effect of the stellar irradiation on the planet structure and evolution. (1) Effect of a core and metal enrichment: dotted line: adiabatic interior, no core, $Z=Z_\odot$; long-dashed line : adiabatic interior, central dunite core
$M_c=6\,\mearth$, metal enrichment $Z_{env}\simeq 6 \times Z_\odot$ in the envelope. Effect of layered convection (same $M_c$ and $Z_{env}$):
solid line: layered convection for $N$=100 layers; dot-dash line (upper panel): $N$=50 layers. The observed values of
 the seven planets with abnormally large radii, HD209458b (triangle), WASP-1b, XO-1b, OGLE-TR-56b, Tres-2,
HAT-P-1b and HD189733b, with masses ranging from 0.5 to 1.3 $M_{Jup}$, are displayed with their most
recent 1-$\sigma$ error bar determinations.}}
\end{figure}

\clearpage

\begin{figure}
\epsscale{0.8}
\plotone{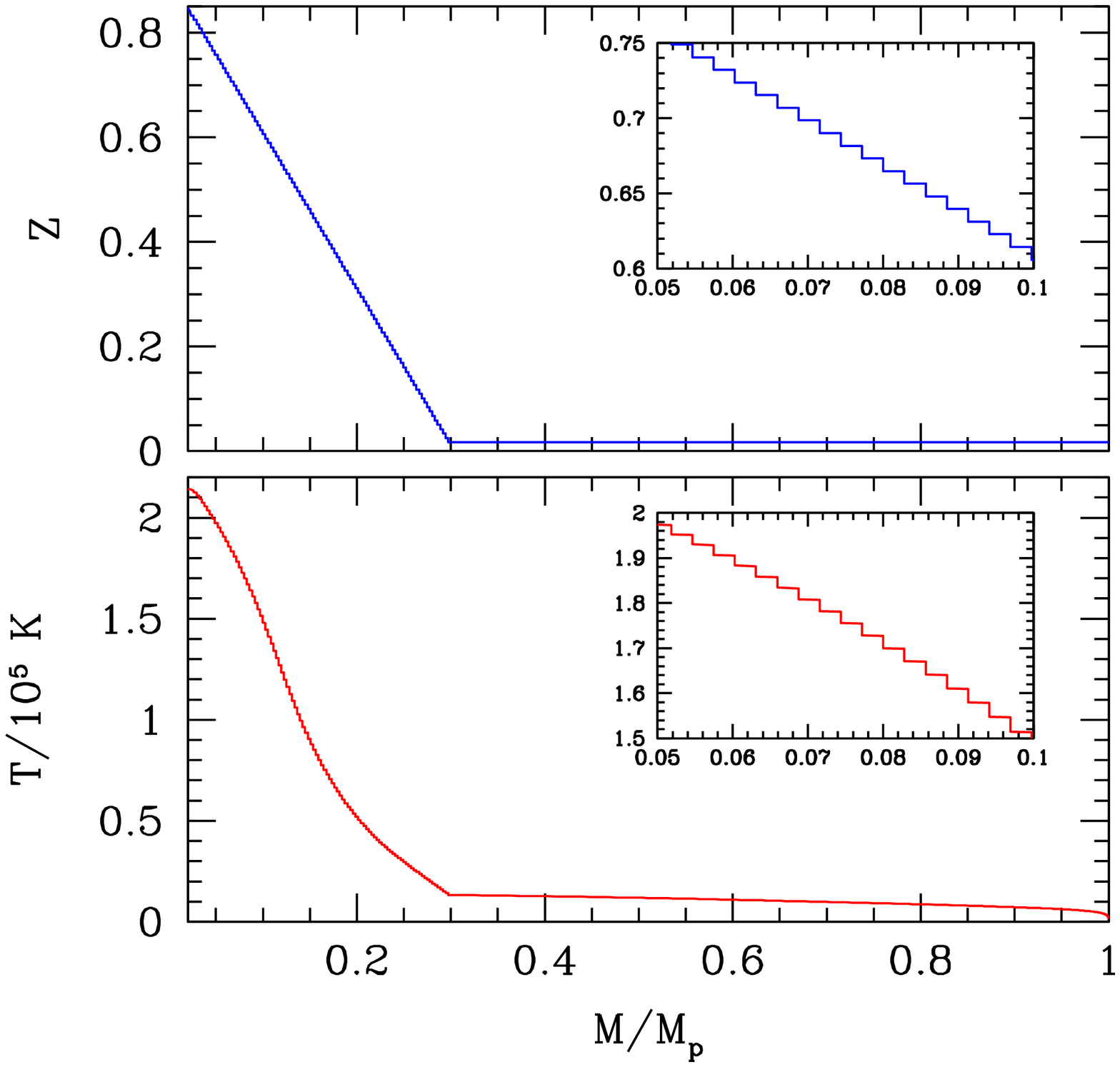} 
\caption{Internal heavy element and thermal profiles at 5 Gyr for the case of 100 convective + diffusive
layers distributed in the inner 30\% by mass of the planet. The global heavy element mass fraction of the planet is $Z=M_Z/M_p=0.13$, including the central core ($M_{core}=2\%\,M_p$). 
Note the step-like T and Z-distributions, as portrayed in the inner subsets.}
\end{figure}


\newpage

\begin{thebibliography}{}
\bibitem[Alibert et al. 2005]{Alibert05}  Alibert, Y., Mordasini,  C., Benz, W. \& Winisdoerffer, C., 2005, \aap,  434, 343 
\bibitem[Baraffe et al. 2005]{B05}  Baraffe, I.  et al., 2005, \aap, 436, 47
\bibitem[Baraffe et al. 2006]{B06}  Baraffe, I., Alibert, Y., Chabrier, G.  \& Benz, W., 2006, \aap,  450, 25
\bibitem[Ferguson et al. 2005]{Ferguson05}  Ferguson et al., J., 2005, \apj,  623, 585
\bibitem[Fernando 1989]{Fernando89}  Fernando, H., 1989, \jfm,   209, 1
\bibitem[Guillot et al. 2004]{Guillot04}  Guillot, T. , Stevenson, D., Hubbard, W.B., Saumon, D., 2004, Jupiter: The Planet, Satellites and Magnetosphere, 
Cambridge University Press 
\bibitem[Iaroslavitz \& Podolak 2007]{IaroslavitzPodolak07} Iaroslavitz, E. \& Podolak, M., 2007, Icarus, 187, 600
\bibitem[Levrard et al. 2007]{Levrard07}  Levrard, B. et al., 2007, \aap,  462, L5
\bibitem[Linden \& Shirtcliffe 1978]{LS78}  Linden, P.F. \& Shirtcliffe, T., 1978, \jfm,  87, 417
\bibitem[Merryfield 1995]{Merryfield95}  Merryfield, W., 1995, \apj,  444, 318
\bibitem[Mizuno 1980]{Mizuno80}  Mizuno, H., 1980,  Prog. Th. Physics,  64, 544
\bibitem[Molemaker \& Dijkstra 1987]{MolemakerDijkstra97}  Molemaker, M.  \& Dijkstra, H., 1997, \jfm,  331, 199 
\bibitem[Pollack et al. 1996]{P96}  Pollack, J. et al., 1996, Icarus,  124, 62 
\bibitem[Potekhin 1999]{P99}  Potekhin, A., 1999, \aap,  351, 787
\bibitem[Rogers et al. 1996]{OPAL} Rogers, F., Swenson, F. \& Iglesias, C., 1996, \apj, 456, 902
\bibitem[Saumon \& Guillot 2004]{SG04} Saumon,  D. \& Guillot, T. , 2004, \apj,  609, 1170
\bibitem[Showman \& Guillot  2002]{SG02}  Showman, A., \& Guillot, T., 2002, \aap,  385, 156
\bibitem[Schmitt 1994]{Schmitt}  Schmitt, R., 1994,  Ann. Rev. Fluid Mech.,  26, 255 
\bibitem[Spruit 1992]{Spruit92} Spruit, H., 1992, \aap,  253, 131 
\bibitem[Stevenson 1979]{Stevenson79}  Stevenson, D., 1979, \mnras,  187, 129
\bibitem[Stevenson 1982]{Stevenson82}  Stevenson, D., 1982,  Planet. Space Sci., 755
\bibitem[Stevenson \& Salpeter 1977a]{SS77} Stevenson, D., Salpeter, E., 1977a, \apjs,  35, 221
\bibitem[Stevenson \& Salpeter 1977b]{SS77b} Stevenson, D., Salpeter, E., 1977b, \apjs, 35, 239 
\end{thebibliography}
\end{document}